\def\beq#1{\begin{equation}\label{#1}}
\def\eeq{\end{equation}}
\def\beqa#1{\begin{eqnarray}\label{#1}}
\def\eeqa{\end{eqnarray}}

\def\fun#1#2{\lower3.6pt\vbox{\baselineskip0pt\lineskip.9pt
        \ialign{$\mathsurround=0pt#1\hfill##\hfil$\crcr#2\crcr\sim\crcr}}}

\def\xi{{{\bf x}^b}}

\def\bfp{\mbox{\bf p}}
\def\bfq{\mbox{\bf q}}

\documentclass[twocolumn,           
               showpacs,            
               preprintnumbers,     
               aps,                 
               prd,                 
               a4paper,             
               superscriptaddress,  
               nofootinbib,         
               tightenlines,        
               floats,floatfix      
               ]{revtex4}

\usepackage{graphicx}
\usepackage{subfigure}
\usepackage{latexsym}
\usepackage{amsmath,amssymb}        
\usepackage[draft=false]{hyperref}
\usepackage{epsfig}

\newcommand{\be}{\begin{equation}}
\newcommand{\ee}{\end{equation}}
\newcommand{\ba}{\begin{eqnarray}}
\newcommand{\ea}{\end{eqnarray}}

\input colordvi

\begin{document}


\title{Planck priors for dark energy surveys}
\author{Pia~Mukherjee}
\affiliation{Astronomy Centre, University of Sussex, Brighton BN1 9QH,
United Kingdom}
\author{Martin~Kunz}
\affiliation{Astronomy Centre, University of Sussex, Brighton BN1 9QH,
United Kingdom; Department of Theoretical Physics, Univ. of Geneva, 
24 Quai E. Ansermet, 1211 Geneve 4, Switzerland}
\author{David~Parkinson}
\affiliation{Astronomy Centre, University of Sussex, Brighton BN1 9QH,
United Kingdom}
\author{Yun~Wang}
\affiliation{Homer L. Dodge Department of Physics \& Astronomy, Univ. 
of Oklahoma, 440 W Brooks St., Norman, OK, USA}
\date{\today}
\pacs{98.80.-k \hfill astro-ph/yymmnnn}
\preprint{astro-ph/yymmnnn}

\begin{abstract}
Although cosmic microwave background (CMB) anisotropy data
alone cannot constrain simultaneously the spatial curvature and the equation of
state of dark energy, CMB data provide a valuable addition to other
experimental results. However computing a full CMB power spectrum with a
Boltzmann code is quite slow; for instance if we want to work with many 
dark energy and/or modified gravity models, or would like to optimize 
experiments where many different configurations need to be tested, it 
is possible to adopt a quicker and more efficient approach.

In this paper we consider the compression of the projected Planck
CMB data into four parameters, $R$ (scaled distance to last scattering
surface), $l_a$ (angular scale of sound horizon at last scattering), 
$\Omega_b h^2$ (baryon density fraction) and $n_s$ (powerlaw index
of primordial matter power spectrum),
all of which can be computed quickly. We show that, although this
compression loses information compared to the full likelihood,
such information loss becomes negligible when more data is added.
We also demonstrate that the method can be used for scalar field
dark energy independently of the parametrisation of the equation of
state, and discuss how this method should be used for other kinds
of dark energy models.
\end{abstract}



\maketitle

\section{Introduction}
Dark energy model building continues to be an active area of research 
\cite{KunzSap,Carroll04,OW04,Cardone05,Kolb05,Caldwell06,KO06,Felice-us,Koi07,Ng07}. While 
current data remain consistent with a cosmological constant explanation for 
dark energy, other possibilities are not yet ruled out, especially if 
theoretical motivation can 
be found to tighten their predictions about the data \cite{Liddle:2006kn}. 
New theoretical ideas thus
may bolster support in favour of an exotic component of matter or a 
modification of gravity beyond some length scale.

On the observational front, recognizing the need for better data, many
future dark energy surveys have been proposed, classified by the
Dark Energy Task Force as stage III and stage IV experiments \cite{DETFdoc}. 
The realisable constraints from these surveys depend sensitively on the 
external or prior information that will be available in the future. A crucial
external data set will come from the Planck satellite, which will place strong
constraints on a range of cosmological parameters. It is therefore important
to include this data for forecasts and optimisations of instrument performance
for the stage III and IV dark energy surveys. This in turn requires a rapid
way to evaluate the predicted Planck likelihood, preferably without the 
necessity to run a Boltzmann code.

Some of us have shown that the information from the WMAP CMB
experiment  \cite{Spergel06}
can be effectively and 
simply incorporated into a likelihood analysis of Type Ia 
supernovae (SN Ia) and baryon acoustic oscillation (BAO) data by including
in the likelihood a term involving WMAP constraints on 
the CMB shift parameter ($R$), the angular scale of the 
sound horizon at last scattering ($l_a$), and the baryon density 
$\Omega_bh^2$, in Gaussian form
together with their full covariance matrix \cite{WM07}. 
The idea being that the calculation of full CMB spectra at each 
parameter point can be avoided, so that a Markov Chain Monte 
Carlo (MCMC) analysis proceeds very quickly. The merit lies in the method 
being independent of the dark energy model used as long as only 
background (or homogeneous) quantities are varied.

In this paper we extend the method to projected Planck data,
which is significantly more accurate than WMAP data. We derive and test this
 simple prescription, compare it to a full likelihood 
analysis of simulated Planck data, and conclude that when such a Planck 
prior is combined with future dark energy surveys useful complementary
 information from the CMB is retained and there is no significant 
information loss. Hence the prescription remains an effective way to 
incorporate constraints from Planck (or Planck priors) in the analyses 
of data from future dark energy surveys. 

\section{Components of the proposed Planck likelihood}
Let us first introduce the parameters that we are proposing to use 
as an effective summary of the information contained in a CMB spectrum:
\be
R \equiv \sqrt{\Omega_m H_0^2} \,r(z_{CMB}), \hskip 0.1in
l_a \equiv \pi r(z_{CMB})/r_s(z_{CMB}),
\label{eq1}
\ee
where $r(z)$ is the comoving distance from the observer to redshift $z$,
and $r_s(z_{CMB})$ is the comoving size of the sound-horizon at decoupling.
We give the details of the formulae used in appendix~\ref{app:formulae}.

In this scheme, $l_a$ describes the peak location through the angular diameter
distance to decoupling and the size of the sound horizon at that time.
If the geometry changes, either due to non-zero curvature or due to a different
equation of state of dark energy, $l_a$ changes in the same way as the 
peak structure. $R$ encodes similar information, but in addition 
contains the matter density which is connected with the peak height.
In a given class of models (for example, quintessence
dark energy), these parameters are ``observables'' relating to 
the shape of the observed CMB spectrum, and constraints on them 
remain the same independent of (the prescription for) the equation of 
state of the dark energy.
Furthermore, $R$ and $l_a$ are very well constrained by WMAP and even 
better by Planck and their likelihoods are almost perfectly Gaussian 
(remaining so under different treatments of dark energy), so that a Gaussian
likelihood term together with the corresponding covariance matrix 
retains almost all
of the information on these derived parameters. With curvature 
held fixed, an even simpler set up using just $R$ sufficed and 
has been used by many authors, including \cite{WM06,WM04}. 

As a caveat we note that if some assumptions regarding the evolution of
perturbations are changed, then the corresponding 
$R$ and $l_a$ constraints and covariance 
matrix will need to be recalculated under each such hypothesis, for 
instance if massive neutrinos were to be included, or even if tensors were 
included in the analysis \cite{CM}. Further $R$ as defined in Eq.~(\ref{eq1})
can be badly constrained and quite useless if the dark energy clusters
as well, e.g. if it has a low sound speed, as in the model discussed in
\cite{deg}. However,
as discussed further below we checked that our constraints are
valid at least for scalar-field dark energy models, independent
of the parametrisation of $w(z)$.

In addition to $R$ and $l_a$ we use the
baryon density $\Omega_bh^2$, and optionally the spectral index of the 
scalar perturbations $n_s$, as these are strongly
correlated with $R$ and $l_a$, which means that we will lose information
if we do not include these correlations.

\section{Simulated data}

Our simulation and treatment of Planck data is as in \cite{Pahud06}. 
We include the temperature and 
polarization (TT, TE, and EE) spectra from three temperature channels with 
specification similar to the HFI channels of frequency 100 GHz, 143 GHz, 
and 217 GHz, and one 143 GHz polarization channel, following the current
Planck documentation,\footnote{
www.rssd.esa.int/index.php?project=PLANCK\&page=perf\underline{~}top}.
The full likelihood is constructed assuming a sky coverage of 0.8. 
We choose a fiducial model close to the WMAP best fit LCDM model:
$\Omega_b h^2 = 0.022$, $\Omega_m h^2 = 0.127$, $h=0.73$, 
$\Omega_k =0$, $w_0=-1$, and $w_a=0$.

For the Baryon Acoustic Oscillation part, we use the experimental 
configuration outlined in the DETF report \cite{DETFdoc}. A Stage 
III spectroscopic experiment would cover 2000 square degrees with a 
redshift range of   $0.5< 1.3$, divided into 4 equally sized 
redshift bins, plus 300 square degrees with $ 2.3 < z < 3.3$. The 
experiment would obtain the spectra of $10^7$ galaxies. This
survey will measure the oscillations in the galaxy power
spectrum, in the tangential direction (measuring $r(z)$), and the 
radial direction (measuring $dr(z)/dz = c/H(z)$, providing a direct
measurement of the Hubble parameter). To estimate the accuracy with 
which the radial and tangential 
oscillations can be measured, we apply these survey parameters to 
the fitting formulae described in \cite{BAOfitting}. These fitting
formula only consider the accuracy with which the oscillations 
themselves can be measured, returning no information about the 
accuracy of the power spectrum measurement (as is done in e.g. 
\cite{SeoEisenstein,Wang06}. This is because the number of possible 
parameters contributing to the nature of the matter power spectrum, such  
as running of the spectral index, massive neutrinos, and non-linear 
bias, make this calculation very assumption dependent. In contrast, 
the positions of the oscillations is very robust with regard to
these extra considerations.

For the Supernovae, we use a Stage III spectroscopic survey
as described in the DETF report \cite{DETFdoc}. We assume a scaled-up 
version of the SNLS survey with 2000 supernovae in the range 
$ 0.1 < z < 1$, with a further 500 supernovae at low redshift. The 
dispersion in observed magnitude is the sum in quadrature of a fixed 
$\sigma_D = 0.12$ with a second piece $\sigma_m$, which
is fixed at $0.02$ up to $z = 0.4$ but then increases up to 0.03 by $z=1$.

\section{Analysis}

The full set of constraints on all parameters including $R$ and $l_a$
are determined through an MCMC based likelihood analysis \cite{LewisBridle} of 
simulated Planck data. Planck will provide much tighter constraints on 
parameters, and its posterior will be significantly better localized in 
parameter space than that of WMAP. The shape of the posterior 
(i.e. parameter correlations) is also found to be quite different 
from WMAP's (further justifying the exercise of determining the best way to 
incorporate constraints from Planck separately from WMAP). 
While $R$ and $l_a$ were almost 
uncorrelated for WMAP data, this is no longer the case for Planck. Tables 1
and 2 show the estimated values and the covariance matrix for $R$, $l_a$, 
$\Omega_bh^2$ and $n_s$. We have included $n_s$
here because it is found to have a correlation with $R$ and $l_a$ and a 
different consideration of BAO data in the future, utilising the 
full shape of the matter power spectrum, might require the inclusion of $n_s$ 
as a parameter. In the analysis presented in this paper, given 
the conservative treatment of the BAO signal, the inclusion of $n_s$ does not 
have a noticeable impact.

The first point to consider and re-test with Planck data is whether 
the constraints on $R$, $l_a$, $\Omega_bh^2$ and $n_s$, and 
their corresponding covariance matrix 
are independent of the dark energy prescription used. In \cite{WM07} we tested 
this for WMAP data for a cosmological constant, constant $w$ and 
$w_0$-$w_a$ models of dark energy, with and without curvature.
Here we test it again for a flat model with a cosmological constant, 
and the $w_0$,$w_a$ model and the kink model for dark energy, 
both with curvature, and for Planck quality data. 
In the $w_0$,$w_a$ model $w_X(z)=w_0+w_a(1-a)$ \cite{Chev01}
which corresponds to $X(z)= a^{-3(1+w_0+w_a)}e^{3w_a(a-1)}$. In the kink model
the equation of state parameter $w_X$ is described by its value today, $w_0$,
its asymptotic value at high redshift, $w_m$, as well by two more parameters
giving the location and speed of the transition from $w_m$ to $w_0$ 
\cite{CKPCB}.
In this case the energy density is derived through a numerical integration
of the continuity equation. We found that there is no significant difference 
in the constraints on $R$, $l_a$, $\Omega_bh^2$ and $n_s$ obtained using 
these different models. See Figure 1.

\begin{figure}
\begin{center}
\epsfig{figure=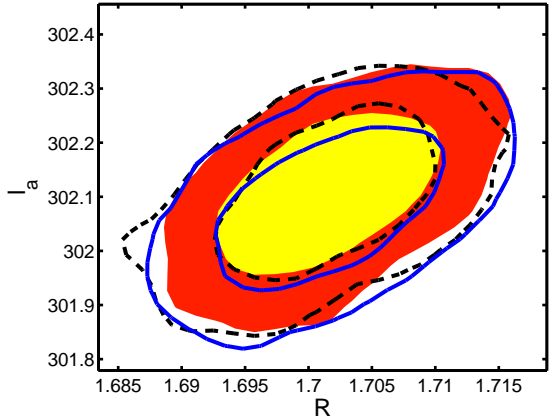,width=8cm}
\caption{This figure shows the projected 68\% and 95\% Planck constraints on $R$ and $l_a$ 
obtained assuming that dark energy were due to a cosmological constant (flat
$\Lambda$CDM, dashed contours), a
 $w_0$,$w_a$ model (shaded contours) and a kink model (solid contours), 
as described in the text.\label{fig1}}
\end{center}
\end{figure}

\begin{figure}
\begin{center}
\epsfig{figure=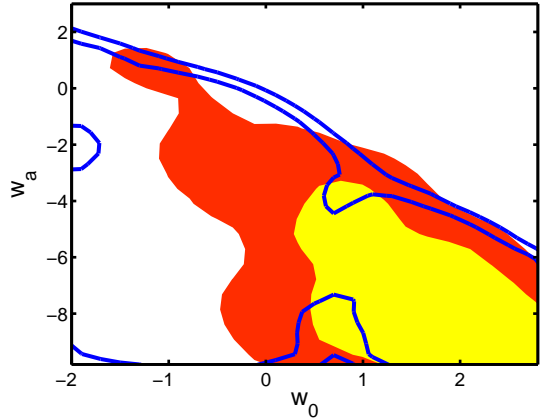,width=8cm}
\caption{This figure shows the Planck projected constraints on $w_0$,$w_a$ obtained using a full likelihood analysis of simulated Planck data (shaded contours) and a simpler and quicker likelihood analysis based on $R$, $l_a$, $\Omega_bh^2$ and $n_s$ (solid contours). Information is thus lost by the simplified analysis.\label{fig2}}
\end{center}
\end{figure}

Let us now test for the amount of information on parameters 
relevant to dark energy that is lost by considering a
likelihood based on $R$, $l_a$, $\Omega_bh^2$ and $n_s$ rather than the full 
CMB spectra. Figure 2 shows $w_0$,$w_a$ contours obtained from a full 
likelihood 
analysis of Planck simulated data (shaded contours) against contours reconstructed 
from the $R$, $l_a$, $\Omega_bh^2$ and $n_s$ likelihood (solid curves). 
We find that even in this limited 2D view there is significant 
information loss: The shaded contours from the full likelihood cover
significantly less area than the open contours from the simpler
likelihood. Due to the strong degeneracies which leave Planck basically
unable to constrain cosmological parameters relating to dark energy and 
curvature on its own, the resulting contours depend strongly on the 
priors used.

It may be useful to note that given the $R$, $l_a$, $\Omega_bh^2$ and $n_s$
likelihood, one can implement a full likelihood analysis under 
different dark energy models more efficiently using  
Hamiltonian Monte Carlo. In this method the 
$R$, $l_a$, $\Omega_bh^2$ and $n_s$ likelihood is used as a guide to or 
an approximation of the true likelihood surface, but at each accepted point
the likelihood is weighted using the full CMB spectra. We discuss this
procedure in more detail in appendix \ref{app:hmc}.

\begin{figure}
{\epsfig{figure=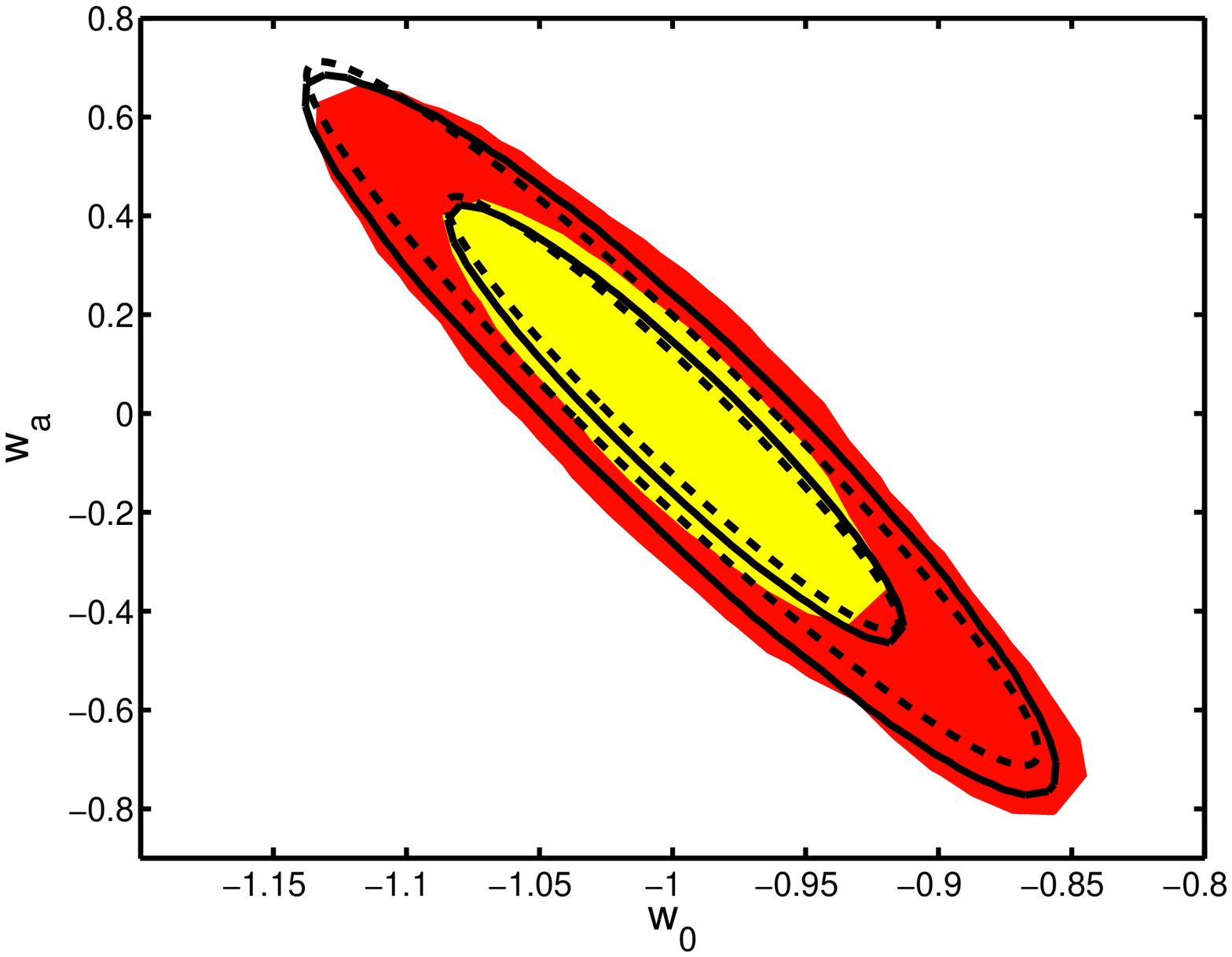,width=8cm}}
{\hspace*{0.5cm}
{\epsfig{figure=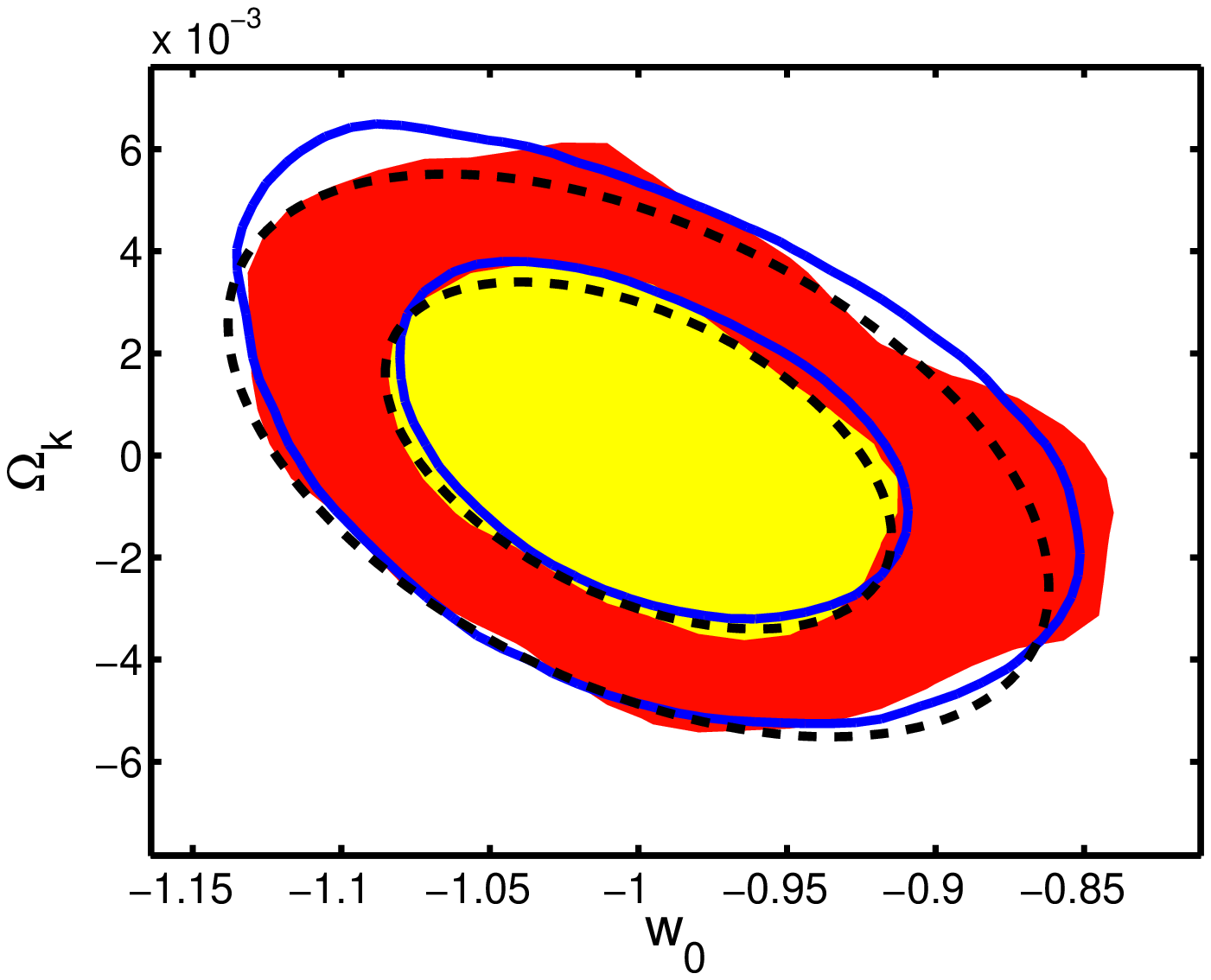,width=8cm}}}
{\hspace*{0.5cm}
{\epsfig{figure=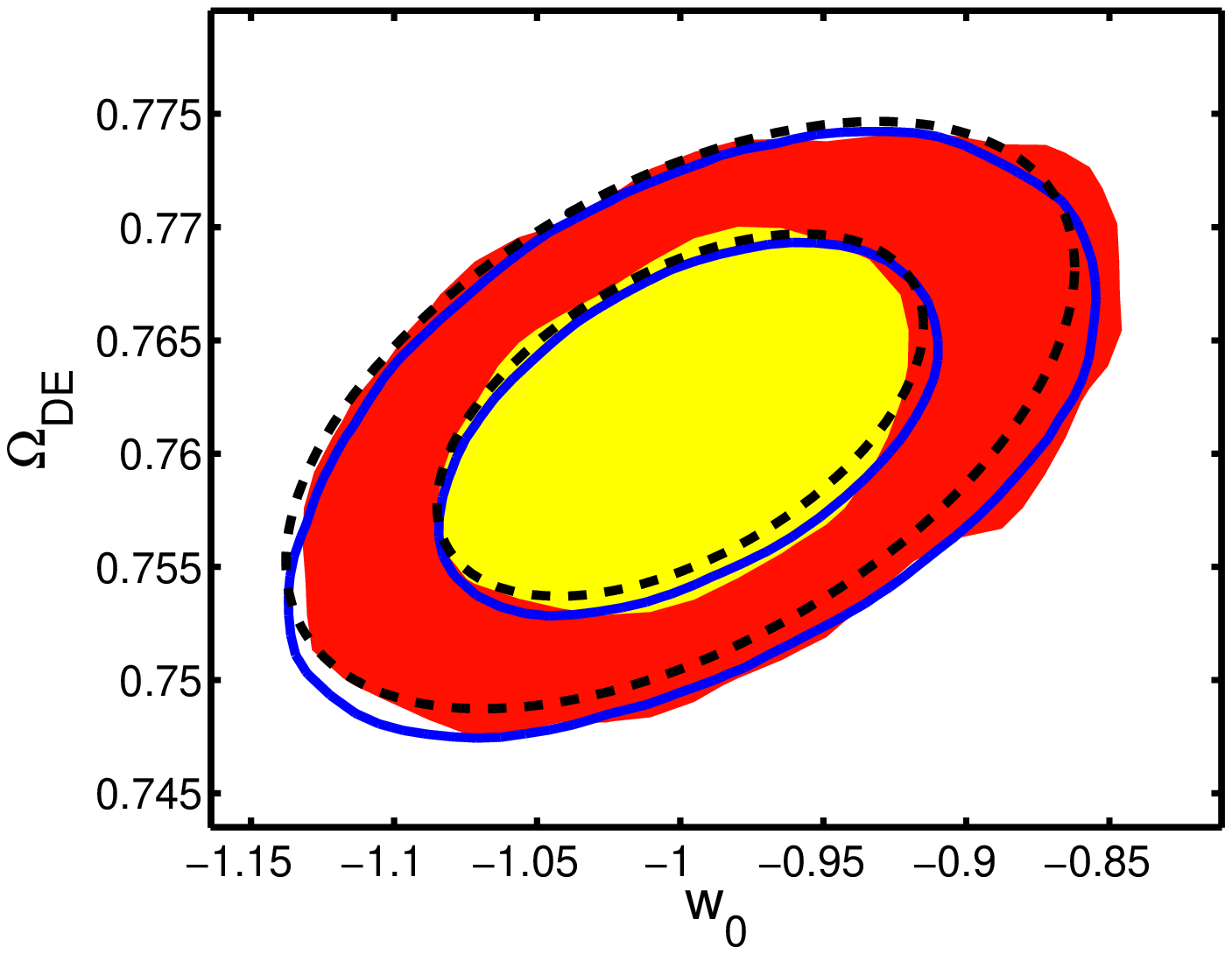,width=8cm}}}
\caption{This figure shows 2D confidence contours obtained using a full likelihood analysis of simulated Planck data in conjunction with stage III SN Ia and BAO data (shaded contours), contours obtained using the simplified $R$, $l_a$, $\Omega_bh^2$ and $n_s$ based likelihood analysis of Planck data together with stage III SN Ia and BAO data (solid contours) and finally a Fisher matrix treatment of all data (dashed contours) as described further in the text.\label{fig3}}
\end{figure}

However the question remains, whether there is still an information loss 
when using the $R$, $l_a$, $\Omega_bh^2$ and $n_s$ likelihood from Planck
when analysing SN Ia and/or BAO data, 
as compared to the full Boltzmann analysis of Planck data, which is much more 
time consuming and so limits our ability to consider many varied 
dark energy models. To address this we compared the outputs from two 
analyses. Firstly we performed a full MCMC run, i.e. including 
the full Planck likelihood and likelihoods from simulated stage III 
SN Ia and BAO surveys. Secondly we perfomed a MCMC analysis  using the 
$R$, $l_a$, $\Omega_bh^2$ and $n_s$ likelihood from Planck together with 
the SN Ia and BAO likelihood. Fig.~\ref{fig3} shows the 
constraints obtained in each case. We conclude that there is effectively no 
information loss in using the $R$, $l_a$, $\Omega_bh^2$ and $n_s$ 
likelihood, in conjuction with the likelihood from a better SN Ia and/or BAO 
experiment, and this condensed data analysis proceeds much faster than 
an analysis involving the full CMB likelihood.

Another way to include a Planck prior in forecasting constraints from a 
future dark energy experiment is to do it via a Planck Fisher matrix. 
Consider the above likelihood analysis in comparison to a Fisher matrix
 treatment. The Planck Fisher 
matrix was obtained from the Planck covariance matrix
of ($R$, $l_a$, $\Omega_b h^2$, $n_S$), with the appropriate 
parameter transformations for compatibility with 
the SN Ia and BAO Fisher matrices. See Appendix B for a description of 
how the Planck Fisher matrix was obtained and Table IV for the 
resulting Planck Fisher matrix. Constraints on 
dark energy parameters obtained in this way are also shown in Fig.{\ref{fig3}}.
Because of the nearly unconstrained directions, the pure Planck Fisher
matrix cannot be inverted, as the range of eigenvalues is larger than
its precision. This can be rectified with weak priors on the parameters
(in which case diagonal entries in the inverse of the Fisher matrix will
reflect those priors), or by adding more data. Figure 3 shows that the
Fisher matrix is valid in spite of its formal problems: the error contours
for Planck + SN-Ia + BAO data agree very well with the others.

\section{Conclusions}
We have found that a Gaussian likelihood based on $R$, $l_a$, 
$\Omega_bh^2$ and $n_s$ effectively summarizes the information in Planck
that is relevant for an analyses of data from SN Ia and BAO
experiments for dark energy parameters under different 
dark energy models.
Therefore a Planck prior can be included in this manner. When used in
conjunction with other data that are more sensitive to dark energy 
such a treatment of Planck data results in no 
information loss as compared to a full analysis, while being much faster.

We provide the full $R$, $l_a$, $\Omega_bh^2$, $n_s$ covariance matrix 
that is required to define such a likelihood from Planck. We also provide 
a Planck Fisher matrix for people who prefer to use the Fisher matrix route 
to forecasting constraints for a future experiment.  
Using such a Planck prior we have obtained the type of constraints 
that may be expected from a stage III SN Ia and BAO survey. 
Of course the prescription can also be used once data from all these 
experiments have actually been obtained (ie. the prescription is not just 
for forecasting).

In the above analysis we found that it was not strictly necessary 
to include $n_s$ given our conservative treatment of BAO data. A 
fuller treatment of BAO data such as one that 
included the shape of the matter power spectrum rather
than the transverse and line of sight distances to the redshifts of the BAO
survey deduced from the BAO scales in the corresponding directions, 
would require the primordial power spectrum parameters including $n_s$ to
be considered a variable in the BAO part of the analysis. For this reason we
have included $n_s$ in our prescription, and marginalized over it in our 
results.

While this work was in progress \cite{CM} considered a likelihood analysis 
involving the locations of the peaks and troughs in the CMB 
spectrum observed by WMAP to constrain dark energy parameters in 
combination with recent BAO data. This offers another way to include 
information from the CMB in a likelihood analysis of BAO and SN Ia data.
It involves fitting formulae for the locations of the extrema presented 
in \cite{DL}. Fitting formulae have been derived to account for certain 
pre-recombination effects that via the 
early ISW effect can effect the position of the first peak relative to the 
higher peaks. In our formalism we would have to recompute the $R$, $l_a$, 
$\Omega_bh^2$ and $n_s$ constraints for each new pre-recombination scenario, 
such as involving a non-zero neutrino mass, involving tensors and/or the 
running of the scalar spectral index, or else include these parameters
in the covariance matrix. On the other hand, our approach is arguably
simpler to implement, and at least as accurate within its domain of 
applicability (since it additionally uses $R$ as an effective measure of 
peak height). In their approach too new fitting formulae would have to 
be derived new effects in different scenarios that haven't been considered 
in the past.

\section*{Acknowledgements}
We thank Pier-Stefano Corasaniti and Julien Larena for interesting discussions.
MK acknowledges partial funding by the Swiss NSF. PM acknowledges the Department 
of Physics and Astronomy, University of Sussex, for support.

\appendix

\section{Detailed description and formulae\label{app:formulae}}

The Planck satellite will deliver data of such a high quality that even
small changes in parameters like the CMB temperature can have a significant
impact. For this reason we summarise here the relevant formulae used in
this paper. Generally they are those used by CAMB.

The comoving distance to a redshift $z$ is given by
\ba
\label{eq:rz}
&&r(z)=cH_0^{-1}\, |\Omega_k|^{-1/2} {\rm sinn}[|\Omega_k|^{1/2}\, \Gamma(z)]\\
&&\Gamma(z)=\int_0^z\frac{dz'}{E(z')}, \hskip 1cm E(z)=H(z)/H_0 \nonumber
\ea
where $\Omega_k=-k/H_0^2$ with $k$ denoting the curvature constant, 
and ${\rm sinn}(x)=\sin(x)$, $x$, $\sinh(x)$ for 
$\Omega_k<0$, $\Omega_k=0$, and $\Omega_k>0$ respectively, and
\ba
E(z)&=&\left[\Omega_m (1+z)^3+\Omega_{\rm rad}(1+z)^4 \right. \\
          &&\left. +\Omega_k(1+z)^2+\Omega_X X(z)\right]^{1/2}
\ea
with $\Omega_X=1-\Omega_m-\Omega_{\rm rad}-\Omega_k$, and the dark energy density
function $X(z) \equiv \rho_X(z)/\rho_X(0)$.

We calculate the distance to decouplingm, $z_{CMB}$, via the fitting 
formula in \cite{HuSugiyama}. 
CAMB \cite{camb} uses the same fitting formula. We note that simply 
using a constant for $z_{CMB}$ results in a shift in the inferred values
of the CMB shift parameters at levels of precision corresponding to Planck.
The comoving sound horizon at recombination is given by
\ba
\label{eq:rs}
r_s(z_{CMB}) &=& \int_0^{t_{CMB}} \frac{c_s\, dt}{a}
=cH_0^{-1}\int_{z_{CMB}}^{\infty} dz\,
\frac{c_s}{E(z)}, \nonumber\\
&=& cH_0^{-1} \int_0^{a_{CMB}} 
\frac{da}{\sqrt{ 3(1+ \overline{R_b}\,a)\, a^4 E^2(z)}},
\ea
where $a$ is the cosmic scale factor, 
$a_{CMB} =1/(1+z_{CMB})$, and
$a^4 E^2(z)=\Omega_{\rm rad}+ \Omega_m a+\Omega_k a^2 +\Omega_X X(z) a^4$.
The radiation density is computed using the Stefan-Boltzmann formula from the
CMB temperature, assuming $3.04$ families of massless neutrini.
The sound speed is $c_s=1/\sqrt{3(1+\overline{R_b}\,a)}$,
with $\overline{R_b}\,a=3\rho_b/(4\rho_\gamma)$,
$\overline{R_b}=31500\Omega_bh^2(T_{CMB}/2.7\,{\rm K})^{-4}$.\footnote{
We used a $T_{CMB}=2.726$, and 
$\overline{R_b}=30000\Omega_bh^2$
as defined in CAMB, noting that precision can be improved by updating 
these definitions.}

\section{Fisher matrix approach}

The Fisher matrix, $F_{\alpha \beta}$, for a set of parameters $\bfp$ can be derived from
the Fisher matrix, $F_{ij}$, for a set of equivalent parameters $\bfq$ as follows \cite{Wang06}
\be
F_{\alpha \beta}= \sum_{ij} \frac{\partial p_i}{\partial q_{\alpha}}\,
F_{ij}\, \frac{\partial p_j}{\partial q_{\beta}}.
\label{eq:Fisherconv}
\ee
The Fisher matrix of $\bfq=(R$, $l_a$, $\omega_b$, $n_S$) is the inverse of
the covariance matrix of $\bfq$ (given in Tables I and II).
Note that the CMB shift parameters $R$ and $l_a$ encode all the information on dark energy
parameters. For any given dark energy model parameterized by the parameter
set $\bfp_X$, the relevant Fisher matrix for $\bfp=(\bfp_X$, $\Omega_{DE}$, $\Omega_k$, 
$\omega_m$, $\omega_b$, $n_S$) can be found using Eq.(\ref{eq:Fisherconv}).
For the case most discussed in the literature, $w_X(z)=w_0+w_a(1-a)$,
$\bfp_X$=($w_0$, $w_a$).

In order to find the Fisher matrix for 
($w_0$, $w_a$, $\Omega_{DE}$, $\Omega_k$, $\omega_m$, $\omega_b$, $n_S$),
the following derivatives are needed:
\ba
&& \frac{\partial R}{\partial w_i}=\frac{\partial \Gamma(z_{CMB})}{\partial w_i}\,
\sqrt{\Omega_m}\, {\mbox{cosn}}\left[|\Omega_k|^{1/2} \Gamma(z_{CMB})\right] \nonumber\\
&& \frac{\partial \ln R}{\partial \Omega_{DE}}= -\frac{1}{2 \Omega_m}+ |\Omega_k|^{1/2}\, \nonumber\\
&& \;\;\;\; \frac{{\mbox{cosn}}\left[|\Omega_k|^{1/2} \Gamma(z_{CMB})\right]}
{{\mbox{sinn}}\left[|\Omega_k|^{1/2} \Gamma(z_{CMB})\right]} 
\frac{\partial \Gamma(z_{CMB})}{\partial\Omega_{DE}}
\nonumber \\
& & \frac{\partial \ln R}{\partial \Omega_k}= -\frac{1}{2 \Omega_m}-\frac{1}{2 \Omega_k}+ |\Omega_k|^{1/2}\, \nonumber \\
&& \;\;\;\; \frac{{\mbox{cosn}}\left[|\Omega_k|^{1/2} \Gamma(z_{CMB})\right]}
{{\mbox{sinn}}\left[|\Omega_k|^{1/2} \Gamma(z_{CMB})\right]} 
\left[\frac{\partial \Gamma(z_{CMB})}{\partial\Omega_k}
+\frac{\Gamma(z_{CMB})}{2\Omega_k}\right]
\nonumber \\
&& \frac{\partial R}{\partial \omega_m}=0, \hskip 1cm
\frac{\partial R}{\partial \omega_b}=0, \hskip 1cm
\frac{\partial R}{\partial n_S}=0\nonumber\\
& & \frac{\partial \ln l_a}{\partial w_i}=|\Omega_k|^{1/2}\, 
\frac{{\mbox{cosn}}\left[|\Omega_k|^{1/2} \Gamma(z_{CMB})\right]}
{{\mbox{sinn}}\left[|\Omega_k|^{1/2}\Gamma(z_{CMB})\right]}\nonumber \\
&& \;\;\;\; \frac{\partial \Gamma(z_{CMB})}{\partial w_i}-
 \frac{\partial \ln [H_0r_s(z_{CMB})]}{\partial w_i}\nonumber\\
&&  \frac{\partial \ln l_a}{\partial \Omega_{DE}}=|\Omega_k|^{1/2}\, 
\frac{{\mbox{cosn}}\left[|\Omega_k|^{1/2} \Gamma(z_{CMB})\right]}
{{\mbox{sinn}}\left[|\Omega_k|^{1/2} \Gamma(z_{CMB})\right]} \nonumber \\
&& \;\;\;\; \frac{\partial \Gamma(z_{CMB})}{\partial \Omega_{DE}}
-\frac{\partial \ln [H_0r_s(z_{CMB})]}{\partial \Omega_{DE}}\nonumber\\
&& \frac{\partial \ln l_a}{\partial \Omega_k}=-\frac{1}{2 \Omega_k}+|\Omega_k|^{1/2}\, 
\frac{{\mbox{cosn}}\left[|\Omega_k|^{1/2} \Gamma(z_{CMB})\right]}
{{\mbox{sinn}}\left[|\Omega_k|^{1/2}\Gamma(z_{CMB})\right]}\nonumber \\
&& \;\;\;\; \left[\frac{\partial \Gamma(z_{CMB})}{\partial \Omega_k}+
\frac{\Gamma(z_{CMB})}{2\Omega_k}\right]
-\frac{\partial \ln [H_0r_s(z_{CMB})]}{\partial \Omega_k}\nonumber\\
& & \frac{\partial \ln l_a}{\partial \omega_m}=-\frac{\partial \ln [H_0r_s(z_{CMB})]}
{\partial \omega_m},\nonumber \\
& & \frac{\partial \ln l_a}{\partial \omega_b}=-\frac{\partial \ln [H_0r_s(z_{CMB})]}
{\partial \omega_b},\nonumber \\
& & \frac{\partial l_a}{\partial n_S}=0 \nonumber\\
& & \frac{\partial \omega_b}{\partial \omega_b}=1, 
\hskip 1cm
\frac{\partial \omega_b}{\partial p_i}=0\,\, (p_i \neq \omega_b)\nonumber\\
& & \frac{\partial n_S}{\partial n_S}=1, 
\hskip 1cm
\frac{\partial n_S}{\partial p_i}=0\,\, (p_i \neq n_S),
\ea
where $w_i=(w_0,w_a)$, and ${\rm cosn}(x)=\cos(x)$, $x$, $\cosh(x)$ for 
$\Omega_k<0$, $\Omega_k=0$, and $\Omega_k>0$ respectively.

Note that in the limit of $\Omega_k=0$, 
\ba
&& \frac{\partial \ln R}{\partial \Omega_k}=
\frac{\partial \ln\Gamma (z_{CMB})}{\partial \Omega_k}
-\frac{1}{2\Omega_m} + \frac{[\Gamma(z_{CMB})]^2}{6}\nonumber\\
&&\frac{\partial \ln l_a}{\partial \Omega_k}=
\frac{\partial \ln\Gamma (z_{CMB})}{\partial \Omega_k}
-\frac{\partial \ln [H_0 r_s(z_{CMB})]}{\partial \Omega_k} + \nonumber \\
&& \;\;\;\; \frac{[\Gamma(z_{CMB})]^2}{6}
\ea

For the fiducial model considered in this paper, the Planck Fisher matrix
for ($w_0$, $w_a$, $\Omega_{DE}$, $\Omega_k$, $\omega_m$, $\omega_b$, $n_S$)
is derived from the Planck covariance matrix of ($R$, $l_a$, $\omega_b$, $n_S$)
given in Table IV.

\section{Using our likelihood for Hamiltonian Monte Carlo\label{app:hmc}}

While most cosmological codes use the standard Metropolis MCMC
algorithm, there are other MC approaches which may provide faster exploration especially
in high dimensions. One example is Hamiltonian Monte Carlo (HMC) \cite{McKay,princeton}
where each parameter $\theta_i$ acquires a partner corresponding to a momentum variable $\pi_i$, and
the log-likelihood is regarded as a potential. The momenta are drawn from a univariate normal
probability distribution and the next step in the MCMC exploration is chosen
based on a Hamiltonian motion in this system, with total energy $E=p^2/2 + \chi^2(\theta)/2$.
At the end the momenta are marginalised over, which provides an ensemble of 
samples of the
remaining parameters which is drawn from the posterior distribution. The main
advantage of the HMC method is that the Hamiltonian motion naturally follows
even complicated shapes of the posterior and in principle every proposal is
accepted. The main drawback is that, in order to follow the trajectory, one
needs to evaluate the gradient of the log-likelihood with respect to the
parameters for dozens of steps, for every single proposal. Each proposal
therefore requires hundred(s) of likelihood evaluations if the gradient
cannot be computed analytically. 

Since we have a reasonable approximation of the likelihood, we can instead
use this approximation to compute the gradients. This means that the motion follows
the $(R,l_a,\omega_b,n_s)$ likelihood and at the end the approximate and the true
likelihood are compared. If the true likelihood is worse than the approximate one,
then we can either assign the ratio as a weight to the new point (importance sampling) or
test for rejection with the usual criterion (rejection sampling). If the true
likelihood is better, then have to assign the ratio as a weight $>1$. 
For this to work we must ensure that the approximate likelihood does not exclude 
parameter regions that the true likelihood would allow.

In our case we find that the procedure works quite well for the case
where the Planck data is combined with the SN Ia and BAO data, since there
the information loss is negligible\footnote{We add additionally $\tau$
and $\ln A_s$ to the set of our parameters, and augment the $(R,l_a,\omega_b,n_s)$ likelihood 
with their 2x2 covariance matrix.}. Indeed, we find about 20\% efficiency (ie
roughly every 5th proposal is accepted, or correspondingly, the average weight
of each point is $0.2$), which is very good, especially since we can move a long
distance and obtain completely uncorrelated samples. Using only the Planck data,
we lose a lot of information, and less than 2\% of the proposals are accepted.
This is still not too bad, considering the complexity of the shape of the posterior,
and that the resulting samples are completely decorrelated. Additionally, burn-in
is very quick for HMC and there is no need for initial runs to determine the optimal
proposal matrix.

\begin{table*}[htb]
\caption{$R$, $l_a$, $\Omega_bh^2$ and $n_s$ estimated from Planck 
simulated data.}
\begin{center}
\begin{tabular}{lll}
\hline
Parameter & mean & rms variance \\
\hline
\hline
& $\Omega_k\neq 0$&\\
\hline
$R$ & 1.7016 & 0.0055\\
$l_a$ & 302.108 & 0.098 \\
$\Omega_b h^2$ & 0.02199 & 0.00017 \\
$n_s$ & 0.9602 & 0.0038\\
 \hline		
\end{tabular}
\end{center}
\end{table*}

\begin{table*}[htb]
\caption{Covariance matrix for $(R, l_a, \Omega_b h^2, n_s)$from Planck.}
\begin{center}
\begin{tabular}{lrrrr}
\hline
\hline
 & $R$ & $l_a$ & $\Omega_b h^2$ & $n_s$ \\
 \hline
 \hline
& & $\Omega_k\neq 0$& &\\
\hline
$R$            & 0.303492E-04  & 0.297688E-03  & $-$0.545532E-06  & $-$0.175976E-04 \\
$l_a$          & 0.297688E-03  & 0.951881E-02  & $-$0.759752E-05  & $-$0.183814E-03 \\
$\Omega_b h^2$ & $-$0.545532E-06 & $-$0.759752E-05 &  0.279464E-07  &  0.238882E-06 \\
$n_s$          & $-$0.175976E-04 & $-$0.183814E-03 &  0.238882E-06  &  0.147219E-04 \\
\hline
\end{tabular}
\end{center}
\end{table*}

\begin{table*}[htb]
\caption{Normalized covariance matrix for $(R, l_a, \Omega_b h^2, n_s)$from Planck.}
\begin{center}
\begin{tabular}{lrrrr}
\hline
\hline
 & $R$ & $l_a$ & $\Omega_b h^2$ & $n_s$ \\
 \hline
 \hline
& & $\Omega_k\neq 0$& &\\
\hline
$R$            & 1.  & 0.553856  & $-$0.592359  & $-$0.832527 \\
$l_a$          & 0.553856  & 1.  & $-$0.465820  & $-$0.491026 \\
$\Omega_b h^2$ & $-$0.592359 & $-$0.465820 &  1.  &  0.372425 \\
$n_s$          & $-$0.832527 & $-$0.491026 & 0.372425 & 1. \\
\hline
\end{tabular}
\end{center}
\end{table*}

\begin{table*}[htb]
\caption{Fisher matrix for ($w_0$, $w_a$, $\Omega_{DE}$, $\Omega_k$, $\omega_m$, 
$\omega_b$, $n_S$) derived from the covariance matrix for
$(R, l_a, \Omega_b h^2, n_s)$ from Planck.}
\begin{center}
\footnotesize{
\begin{tabular}{lrrrrrrr}
\hline
\hline
 & $w_0$ & $w_a$ & $\Omega_{DE}$ & $\Omega_k$ & $\omega_m$ & $\omega_b$ & $n_S$ \\
 \hline
 \hline
$w_0$ &  .172276E+06 &  .490320E+05 &  .674392E+06 & $-$.208974E+07 &  .325219E+07 & $-$.790504E+07 & $-$.549427E+05 \\
$w_a$ &  .490320E+05 & .139551E+05 & .191940E+06& $-$.594767E+06 & .925615E+06 &$-$.224987E+07 &$-$.156374E+05\\
$\Omega_{DE}$ & .674392E+06 & .191940E+06 & .263997E+07& $-$.818048E+07 & .127310E+08 &-.309450E+08 &$-$.215078E+06\\
$\Omega_k$ & $-$.208974E+07 & $-$.594767E+06 & $-$.818048E+07 &  .253489E+08 & $-$.394501E+08 &  .958892E+08 &  .666335E+06\\
$\omega_m$ &  .325219E+07 &  .925615E+06 &  .127310E+08 & $-$.394501E+08 &  .633564E+08 & $-$.147973E+09 & $-$.501247E+06\\
$\omega_b$ & $-$.790504E+07 &$-$.224987E+07 &$-$.309450E+08 & .958892E+08 &$-$.147973E+09  &.405079E+09 & .219009E+07\\
$n_S$ & $-$.549427E+05 & $-$.156374E+05 & $-$.215078E+06  & .666335E+06 & $-$.501247E+06  & .219009E+07  & .242767E+06\\

\hline
\end{tabular}}
\end{center}
\end{table*}

\end{document}